\documentstyle[12pt%,leng
]{article}
\begin{document}
\hfill{UBCTP 92-009}

\begin{center}
{\large\sc{\bf Quarks as Topological Defects or}}

{\large\sc{\bf What is Confined Inside a Hadron?}}
\baselineskip=12pt
\vspace{15pt}

A. Kovner$^{\dagger\$ *}$ and B. Rosenstein$^{ \& **}$
\vspace{10pt}

 Physics Department, University of British Columbia\\
Vancouver, B.C., Canada, V6T 1Z1\\
\vspace{5pt}

\end{center}
\begin{abstract}
We present a picture of confinement based on representation of quarks as
pointlike topological defects. The topological charge carried by quarks
and confined in hadrons is explicitly constructed in terms of Yang - Mills
variables. In 2+1 dimensions we are able to construct a local
complex scalar field $V(x)$, in terms of which
the topological charge is $Q=-\frac{i}{4\pi}
\int d^2x \epsilon_{ij}\partial_i(V^*\partial_i V
-c.c.)$. The VEV of the field $V$ in the confining phase is nonzero and the
charge is the winding number
corresponding to homotopy group $\pi_1(S^1)$.
Qurks carry the charge $Q$ and therefore are topological solitons. The phase
rotation of $V$ is generated by the operator of magnetic flux. Unlike in
$QED$, the $U(1)$ magnetic flux is explicitly broken
by the monopoles. This results in formation of a string between a quark and
an antiquark.
The effective Lagrangian for $V$ is derived in models with adjoint and
fundamental
quarks. Baryon is described as a ${\bf Y}$ - shaped configuration of
strings. In 3+1 dimensions the explicit expression for $V$ and therefore a
detailed picture is not available.
However, assuming the validity of the same mechanism we
point out several interesting qualitative consequences. We argue
that in the Georgi - Glashow
model or any grand unified model the photon (in the Higgs phase) should have a
small
nonperturbative mass. Moreover, $W^\pm$ should be confined although with small
string tension.
\end{abstract}

%	  \newline
$^\dagger$ NSERC international postdoctoral fellow.
	  \newline
$^\$ $Address after September 1, 1992: Theory Division, T-8 MS B285, Los Alamos
National
Laboratory, Los Alamos, NM 87545.
\newline
$^*$KOVNER@PHYSICS.UBC.CA
	  \newline
$^\& $Address after October 1, 1992: Institute of Physics, Academia Sinica,
Taipei 11529,
Taiwan.
\newline
$^{**}$BARUCH@PHYSICS.UBC.CA\newline

\vfill
\baselineskip=24pt
\pagebreak

\section{Introduction.}

\par
Understanding of confinement in QCD is still one of the major
challenges of particle physics.
Confinement is a commonly held intuitive
notion. It eludes, however, precise definition in the framework of QCD.

In the phenomenological quark model \cite{close} we certainly do know what
is meant by ``confinement''. The global color charges are confined, so that
all hadrons are the $SU_c(3)$ singlets.
However although the quark model captures some important physics,
it is not derived
directly from QCD. The global part of the $SU(3)$ gauge group is not a
global symmetry of QCD Hamiltonian. There is no $SU(3)$ symmetry related to the
gauge group in the same way the electric charge $U_e(1)$ symmetry is related
to the $U(1)$ gauge group in QED.

In QED the global $U_e(1)$
symmetry generator is $Q=\int d^d x \psi^\dagger \psi$.
The symmetry transformation rotates the phase of charged fields: $\psi(x)
\rightarrow e^{i\chi}\psi(x)$, but does not act on vector potential $A_\mu(x)$.
This by no means coincides with the global part of the gauge group. The
latter is generated by the Coulomb constraint
\begin{equation}
C(x)\equiv e\psi^\dagger(x) \psi(x) - \partial_i E_i(x)
\label{coulomb}
\end{equation}
and acts not only on $\psi$ but also on $A_\mu$ at spatial infinity.

In QED the first term in the Coulomb constraint, the electric
charge density, $\psi^\dagger \psi$
is itself a gauge invariant quantity. Therefore, the electric charge,
defined as its integral, is an operator acting nontrivially
on physical Hilbert space.
There is no analogous construction in QCD. The Coulomb constraint in QCD is
\begin{equation}
C^a \equiv e\psi^\dagger \lambda^a \psi - \partial_i E_i^a + ef^{abc}A_i^bE_i^c
=0
\label{coulombqcd}
\end{equation}
A transformation which one is tempted to
identify with global color symmetry is generated by the first and the third
terms in
$C^a$:
\begin{equation}
Q^a=e\int dx (\psi^\dagger \lambda^a \psi + f^{abc}A_i^bE_i^c)
\label{colorqcd}
\end{equation}
However, it is easy to see, that the matrix element of $Q^a$
between any two physical states (i.e. states that satisfy the Coulomb
constraint
eq.(\ref{coulombqcd})) vanishes. Indeed, $Q^a$ transform as an adjoint
representation of the
global part of the gauge $SU(3)$
group (generated by $\int C^a$). According to Wigner - Eckart theorem,
$<\alpha|Q^a|
\beta>=0$
when $|\alpha>$ and $|\beta>$
are singlets, in distinction to the Abelian case where this problem
does not arise.

Since global $SU_c(3)$ cannot be given a satisfactory definition, the
immediate question is: what is confined inside hadrons?

The quark fields that appear in the QCD Lagrangian are not gauge invariant
objects and
the $SU(3)$ charge that they carry is nonzero only on nonphysical states in the
``large''
Hilbert space of a gauge theory. On the other hand, one would like to think
about
constituent quarks as about strongly interacting, but real particles.
Intuitively one
would like to associate some kind of a conserved ``quark number'' $Q$ with the
constituents \footnote{The connection between these constituent quarks and the
bare quarks of perturbative QCD is not obvious. The constituents presumably
exist as particles at intermediate scales. At large distances they are
confined. At small distances, since nonabelian theory is asymptotically free,
the "bare" quarks look particle like. The constituent "quark number"
we have in mind does not have therefore much to do with the "global color"
which appears as  a perturbative symmetry of bare quarks in the extreme
ultraviolet limit. Throughout the paper we reserve the term "quarks" for these
constituent objects. Terms like "adjoint quarks" should be understood as
constituent quarks in a theory with matter fields in adjoint representation.
Since we will not be discussing the ultraviolet limit, we will not be dealing
with the bare quarks at all.}. The notion of confinement then can be given a
precise meaning: objects
that carry the charge $Q$ interact linearly and can never be separated. This is
precisely what is meant by confinement in compact QED with $Q$ beeing the
electric
charge.

In QCD, it is said sometimes, that the triality is the confined quantum
number. Triality is a discrete physical global symmetry of YM theories with
matter (quarks) in
fundamental representation. This criterion is however totally inadequate
to account for confinement of adjoint quarks.
The mechanism of confinement is presumably the same for fundamental and
adjoint quarks \cite{greensite,smit}. Thus one has
to look for another charge which is confined.

In this paper, we point out the existence of such
a charge $Q$ in nonabelian gauge theories (see eq.(\ref{qqcd}) in the
following). This $U(1)$ charge is topological, in a sense
that the corresponding current is ``trivially'' conserved
(as also is the case in QED
\footnote{In QED the current $J_\mu$
is a curl of a local gauge invariant quantity $\tilde F_{\mu\nu}\equiv \frac
{1}{2}
\epsilon_{\mu\nu\rho\sigma} F^{\rho\sigma}$. Physically, this means that
electric charge
can be measured performing local measurements at spatial infinity
(Gauss law).}).
The charge $Q$ is defined in any dimensionality. In 2+1 dimensions
one can go further.
Q can be represented as the winding number of a gauge invariant local complex
field $V$.
When the (low energy) theory is rewritten in terms of this vortex field $V$,
the (gauge invariant) quarks are identified as topological solitons.
An explicit topological representation of $Q$ allows one to understand
confinement of quarks as linear confinement of topological defects.
We are able to derive explicitly the effective Lagrangian in the region of
parameter space where perturbatively a nonabelian gauge theory is in a
partially broken Higgs phase. Nonperturbatively it is known that the
confinement and Higgs phases are analytically connected and confinement is the
property of both. Although it is not a priori clear that the confined
constituents are the same in both phases it is likely to be the case. We will
suggest how the effective dual Lagrangian is modified in the confinement
regime.
The main quantitative difference between the picture of confinement in the
Higgs and confinement regimes is that in the former case the string tension is
associated with a mass of a pseudoscalar particle (the massive photon) while in
the latter case with the mass of the lightest scalar glueball.
The picture of quarks as topological defects in 3+1 dimensions is still
lacking. We
will briefly return to this point in section 5.

In this paper we investigate properties of $Q$ in 2+1 dimensional nonabelian
gauge theories. The charge $Q$ is analogous
in many respects to the electric charge in QED.
It was shown in \cite{npb} that in QED there exists a local gauge
invariant scalar complex field, $V$, in terms of which the
electric current has a
standard form of the topological
current associated
with the homotopy group $\pi_1(S^1)$:
\begin{equation}
J_\mu=-\frac{i}{4\pi}\epsilon_{\mu\nu\lambda}
\partial^\nu(V^*\partial^\lambda V - c.c.)
\label{J}
\end{equation}
$V(x)$ has a physical meaning in scalar QED as a field interpolating
Nielsen-Olesen vortices in the Higgs phase. In the Coulomb phase $V$
acquires a nonvanishing expectation value. The manifold of
degenerate  vacua is $S^1$. The degeneracy is due to spontaneous breakdown of
the
magnetic flux symmetry $U_\Phi (1)$ generated by\footnote{The
massless photon is the Goldstone boson that corresponds to spontaneous
breakdown of $U_\Phi(1)$.} $\Phi=\int d^2 x B(x)$
 \cite{npb}.
The electric charge has the form
\begin{equation}
Q=-\frac{i}{4\pi}\int \epsilon_{ij}\partial_i(V^*\partial_j V-c.c.)
\label{q}
\end{equation}
and is the winding number of the phase of $V(x)$.
A typical configuration of $V(x)$ which represents a charged particle located
at point $x$ is a ``hedgehog'' centered at $x$: $V(y)=e^{i\Theta(y-x)}$.

In 2+1 dimensional nonabelian gauge theories considered in this paper
a local vortex field $V$ was constructed
by t'Hooft \cite{thooft1}. We will show that the topological charge
$Q$ carried by quarks, when expressed in terms of this field
has precisely the form of eq.(\ref{q}).

A simple classical picture of linear confinement emerges when we consider an
effective
Lagrangian for the field $V$.
In the models we consider this dual Lagrangian is derived explicitly
in the dilute monopole gas
approximation. This dual Lagrangian is completely gauge invariant, and the
interpretation
therefore is given directly in terms of physical observables.
The operator $V$ has a nonvanishing VEV as in QED.
The main difference between abelian and nonabelian gauge theories is that
the $U(1)$ vacuum degeneracy of QED is lifted in QCD.
The reason is that the continuous flux symmetry in QCD is explicitly broken.
\footnote{When fundamental
charges are absent,
the topological t'Hooft's $Z_N$ subgroup of $U_\Phi (1)$ remains
intact. These models may have another phase in which $Z_N$ is unbroken
and $<V>=0$. This is the Higgs phase in which $Q$ is screened.
When fundamental matter fields are present, t'Hooft's $Z_N$ symmetry is also
broken explicitely. Nevertheless the vortex operator $V$ exists as a local
eigenoperator of $\Phi$ and dual description is still useful.}
The monopole
contribution to the dual Lagrangian is a symmetry breaking term of the type
$V^*+V$.
In the presence of such a term the
topological solitons (quarks) are linearly confined. The reason is
that now the theory does not have a continuum of degenerate vacua and the field
$V$ at
infinity must be as close to its VEV as possible. A topologically nontrivial
configuration must contain a wind of the phase of V. It is energetically
favorable
to perform this wind inside a strip of finite width (Fig. 1)
rather than create a hedgehog - like
configuration where $V\ne <V>$ everywhere in space. The formation of
a string which carries a finite
energy density per unit length leads to linear confinement.
A soliton and an antisoliton are therefore bound by linear potential
into a meson.
The mechanism of confinement is the same as for
linear confinement of defects in condensed matter systems \cite{mermin}.
The definition of $V$ and the derivation of the effective Lagrangian in the
Georgi-Glashow model and its $SU(N)$ analog is given in Sections 2 and 3.

In a theory with fundamental matter the spectrum also contains baryons.
The dual
Lagrangian now
is modified to accommodate the symmetry of baryon number.
This is discussed in Section 4.
 It turns out that
two kinds of solitons are prominent. One type carries one unit of topological
charge $Q$ and a baryon number $1/N$. This is the constituent quark. Another
type carries $N$ units of $Q$ and no baryon number. It is useful to think
about it as a ``constituent gluon''. A baryon consists of $N$
constituent quarks and one gluon. For $SU(3)$ this
gives the {\bf Y}-shaped string picture of a baryon \cite{thooft2}.
The confinement picture of gauge theories with fundamental matter is also
very straightforward and simple in the dual description.

In Section 5 we contrast our approach with that of t'Hooft \cite{thooft1}
and briefly discuss some implications of these ideas for four dimensional
gauge theories.

\section{Topological charge in the Georgi-Glashow model.}

We start with Georgi - Glashow model in 2+1 dimensions for which
the phenomenon of confinement
is qualitatively and even quantitatively understood \cite{polyakov1}.
The Georgi - Glashow model does not contain matter fields in fundamental
representation
and in this sence is different from QCD. However, as mentioned in the
introduction
the quark confinement mechanism that we discuss does not differ qualitatively
for adjoint
and fundamental quarks and therefore it is the simplest model where this
mechanism
can be seen in action.

The Lagrangian of the model is
\begin{equation}
{\cal L}=-\frac{1}{4}F^a_{\mu\nu}F^{a\mu\nu}+
\frac{1}{2}({\cal D}_\mu^{ab}\phi^b)^2+\mu^2\phi^2-\lambda(\phi^2)^2
\label{lgg}
\end{equation}
where
\begin{equation}
{\cal D}_\mu^{ab}\phi^b=\partial_\mu\phi^a-ief^{abc}A_\mu^b\phi^c
\end{equation}
In classical approximation, the model has two phases. For $\mu^2>0$
the gauge symmetry is
unbroken and the standard confinement phenomenon is therefore expected. For
$\mu^2<0$
the $SU(2)$ gauge symmetry is broken down to $U(1)$ and the Higgs mechanism
takes place. Two gauge bosons, $W^\pm$, acquire a mass, while the third,
the ``photon'',
remains massless.

This classical description is only partially correct.
The two ``phases''
are connected analytically \cite{fradkin}. As a result in the continuum
limit there is only one phase. This phase
is reminiscent of the Higgs phase
in that there are electrically charged $W^\pm$ constituents and photons.
However $W^\pm$ are linearly confined and
the photon acquires a nonperturbatively
small mass. There is no qualitative difference between this ``phase'' and what
one
would call the ``confinement phase''. For example, one can show directly in the
continuum
limit \cite{polyakov1} that the fundamental Wilson loop has the area law
behaviour,
which is the unambiguous sign of confinement in a theory with only adjoint
matter.
The confinement of $W^\pm$ from the ``confinement phase'' point of view is just
the confinement of the adjoint quarks, since the longitudinal components of the
massive
vector fields $W^\pm$ which appear in perturbative calculations, are precisely
the
charged degrees of freedom of the quark fields $\phi^a$. In the
remainder of this section therefore, we
do not distinguish between the confinement of $W^\pm$ and the confinement of
quarks.

The model is useful for our purposes since availability of the dilute
monopole gas
approximation in the classically ``gauge broken'' phase makes semiclassical
description of confinement possible \cite{polyakov1}.
In this section we will develop the picture of confinement described in the
introduction
using these semiclassical methods in the Higgs - like region of the parameter
space of
the model. Our hope is, that since there is no phase transition between the
Higgs -like
and the confinement - like regions of the parameter space, the picture
qualitatively
remains correct everywhere. If this is the case, our approach, although it does
not
provide a direct calculational method in the confinement - like region, does
provide a clear qualitative understanding of the confinement phenomenon which
is
the main goal of this paper. We will return to discussion of the confinement
region at the end of this section.

The electromagnetic properties of the Georgi - Glashow model
are similar to those of Polyakov's compact QED \cite{polyakov2}.
In compact QED it is clear that the confined charge is just the
electric charge. Since in the Georgi - Glashow model
the mechanism of confinement is the same, the confined charge should
be a $SU(2)$ gauge invariant generalization of the electric charge of compact
QED.
The generalization which allows the topological
interpretation is
\begin{equation}
J^\mu=\frac{1}{e}\epsilon^{\mu\nu\lambda}\partial_\nu (\tilde
F^a_\lambda\hat\phi^a)
\label{jqcd}
\end{equation}
\begin{equation}
Q=\int d^2 x J_0(x)
\label{qqcd}
\end{equation}
where $\hat\phi^a\equiv{\phi^a\over |\phi|}$.
It is easy to see, that
$W^\pm$ indeed carry the topological charge $Q$. In the vacuum $<\hat\phi^1>=
<\hat\phi^2>=0,
<\hat\phi^3>=1$ (in the unitary gauge). Using the Gauss law
eq.(\ref{coulombqcd})
one identifies $Q$ as $Q^3$ of eq.(\ref{colorqcd}).
As in QED the current is a curl of local gauge invariant dual
field strength $\tilde f_\mu$.
\begin{equation}
\tilde f_\mu= \tilde F^a_\mu\hat\phi^a
\label{fqcd}
\end{equation}
We would like now to
express $Q$ manifestly as a topological charge as in eq.(\ref{q}).

Let us now construct explicitly an operator $V(x)$ in the Hamiltonian
formalism.
In QED$_3$ the analogous operator has been found in \cite{npb}.
\begin{equation}
V(x)=\exp {i\over e}\int d^2y[\epsilon_{ij}{(x-y)_j\over (x-y)^2}
E_i(y)+e\Theta(x-y)J_0(y)]
\label{VQED}
\end{equation}
where $\Theta(x)=arctan\frac{x_2}{x_1}$ is the polar angle.
The crucial property of $V$ which ensures that the zeroth component of
eq.(\ref{J})
is satisfied is the term $\int d^2y \Theta(x-y)J_0(y)$ in the exponent.
Clearly the same term must be present in the definition of $V(x)$ in nonabelian
case.
However this operator by itself is nonlocal. In QED, the
second term in eq.(\ref{VQED}) completed $V$ to
the operator of
a {\it singular} gauge transformation with gauge
 parameter
\begin{equation}
\chi(y)={1\over e}\Theta(x-y)
\end{equation}
This was necessary for $V(x)$ to be a local scalar field.
Since $V(x)$ is an operator of a singular gauge transformation, it does not act
on any local gauge invariant variable, except possibly at the point $x$
\footnote{Note that unlike operators generating regular
gauge transformations, in Hamiltonian formalism, the generator of singular
gauge transformations {\it does act} nontrivially on physical states.}.
Euclidean Green's
functions $<TV(x_1)...V(x_n)>$ in the path integral formalism are given by
the partition function of the model in the presence of magnetic monopoles
of magnetic charge $2\pi/e$ at the points $x_1,...,x_n$ \cite{npb}. Since the
monopoles
satisfy the Dirac quantization condition, the Dirac strings are invisible
and the correlators are Lorentz covariant.

In the nonabelian case the analogous operator of singular gauge transformation
is
\begin{equation}
V(x)=\exp {i\over e}\int d^2y[\epsilon_{ij}{(x-y)_j\over (x-y)^2}
\hat{\phi}^a(y)E^a_i(y)+e\Theta(x-y)J_0(y)]
\label{VQCD}
\end{equation}
The gauge function is now itself field dependent
\begin{equation}
\chi^a(y)={1\over e}\Theta(x-y)\hat\phi^a(\vec y)
\end{equation}
which ensures gauge invariance of $V$.
This is the explicit gauge invariant form of t'Hooft's ``disorder parameter''
\cite{thooft1}. Its locality and Lorentz covariance follow by the same argument
as in the
abelian case.
Since $V(x)$ is a scalar field, not only zeroth but also the spatial components
of
eq.(\ref{J}) are satisfied.
Therefore $Q$ is the topological charge which counts windings of the phase of
$V(x)$.

Our aim now is to write down the low energy effective Lagrangian for the
field $V(x)$.
We will derive it using dilute monopole gas approximation.
The structure of the effective Lagrangian is determined essentially
by symmetries
of the theory. In QED, apart from obvious N\"other
symmetries, a very important
symmetry for construction of effective dual Lagrangian has been
the magnetic flux symmetry. It is generated by the magnetic flux through
the plane $\Phi=\int d^2x B$. The conservation of the corresponding
current $\tilde F_\mu$ is just the homogeneous
Maxwell equation: $\partial^\mu \tilde F_\mu=0$.
In the dual Lagrangian this symmetry is represented linearly.
The reason is that the operator $V$ is an eigenoperator
 of the magnetic field:
\begin{equation}
[V(x),B(y)]=-{2\pi \over e}V(x)\delta^2(x-y)
\label{come}
\end{equation}
As in QED, it is easy to check that $V(x)$ eq.(\ref{VQCD})
is a local eigenoperator of
abelian magnetic field $b(x)\equiv\hat\phi^a \tilde F^a_0$.
\begin{equation}
[V(x),b(y)]=-{2\pi \over e}V(x)\delta^2(x-y)
\label{com}
\end{equation}

In the nonabelian case the flux current $\tilde f_\mu$ defined in
eq.(\ref{fqcd}) is no longer (classically) conserved \cite{thooft1}.
However the following modified current for which $V$ is still an eigenoperator,
is classically conserved:
\begin{equation}
\tilde F^\mu=\tilde f^\mu-\frac{1}{e}\epsilon^{\mu\nu\lambda}\epsilon^{abc}
\hat\phi_a ({\cal D}_\nu \hat\phi)^b({\cal D}_\lambda \hat\phi)^c
\label{F}
\end{equation}
The conservation of the magnetic flux $\Phi\equiv\int d^2x \tilde F_0(x)$
is valid classically but does not
survive quantization. The theory contains Euclidean solutions, the t'Hooft -
Polyakov monopoles.
The elementary monopole is a euclidean configuration with the following
asymptotics
\begin{equation}
\hat\phi^a(x)=\hat r^a \ ,\ \ \ \tilde F^a_\mu(x)=\frac{1}{e}\frac{\hat r^a\hat
r_\mu}{r^2}
\end{equation}
For this solution
\begin{equation}
\partial_\mu \tilde F_\mu=\frac{4\pi}{e}\delta^3(x)
\end{equation}
and
\begin{equation}
\Phi(t=\infty)-\Phi(t=-\infty)=\frac {4\pi}{e}
\label{flux}
\end{equation}
Quantum mechanically this means that once these configurations are taken
into account in the path integral, the magnetic flux is not conserved any more.
However, since only configurations with an integer number of monopoles
have finite action, the total flux can only change by an integer multiple of
$4\pi/e$. Therefore the discrete flux transformations
\begin{equation}
U_k\equiv e^{{ie\over 2} k\Phi}
\label{U}
\end{equation}
are still  symmetries for integer $k$.
\begin{equation}
U_k(t)U^{-1}_k(t')=e^{2\pi i km}=1
\end{equation}
As is seen from eqs.(\ref{com}) and (\ref{U}), not all of them however
act independently on physical Hilbert space.
Any one of
them is equivalent either to the unit transformation or to $U_{k=1}$ since
 on the physical states the
flux takes only integer values (in units of ${2\pi/e}$).
Therefore the dual Lagrangian should be symmetric under $V\rightarrow -V$
\footnote{An independent argument for existence of the residual
$Z_2$ flux symmetry was given by t'Hooft \cite{thooft1}.}.

The effective Lagrangian of the Georgi - Glashow model can be obtained
directly using dilute monopole gas approximation.
Polyakov \cite{polyakov1} introduced the dual (electric) potential $\eta$ via
\begin{equation}
\Box \eta = \frac {8\pi}{e}\rho_m
\label{detagg}
\end{equation}
where $\rho_m(x)=\frac {4\pi}{e}n(x)$ and $n(x)$ is the monopole number
density.
The effective Lagrangian in terms of $\eta$ is
\begin{equation}
{\cal L}=\frac{e^2}{32\pi^2}[(\partial_\mu\eta )^2+
M^2{\rm cos}\eta]
\label{letagg}
\end{equation}
where
 and $M^2\sim \frac{M_W ^{7/2}}{e^3}\exp (-\frac{4\pi}{e^2}
M_W)$ where $M_W$ is the mass of the gauge boson.

The effective Lagrangian eq.(\ref{letagg}) can be rewritten in terms of the
vortex field
$V(x)$ defined in eq.(\ref{VQCD}).
The flux anomaly equation
\begin{equation}
\partial_\mu \tilde F^\mu = \rho_m
\label{ae}
\end{equation}
and  eq.(\ref{detagg}) imply
that in the approximation in which eq.(\ref{letagg}) is valid
\begin{equation}
\tilde F_\mu=\frac{e}{8\pi}\partial_\mu\eta
\label{f1}
\end{equation}
{}From eq.(\ref{letagg}) it then follows
\begin{equation}
[\eta(x),B(y)]=\frac {4\pi i}{e}\delta^2(x-y)
\label{cometa}
\end{equation}
or
\begin{equation}
[e^{i\eta(x)/2},B(y)]=-\frac {2\pi}{e}e^{i\eta(x)/2}\delta^2(x-y)
\end{equation}
The operator $e^{i\eta/2}$ is local, scalar, unitary eigenoperator of the
magnetic
field $B(x)$ and is therefore identical with $V(x)$ of eq.(\ref{VQCD}).

We therefore replace $e^{i\eta(x)/2}$ by $V(x)$ and obtain
\begin{equation}
{\cal L}=\frac{e^2}{8\pi^2}\partial_\mu V^*\partial^\mu
V+\frac{e^2M^2}{64\pi^2}
(V^2 +V^{*2})
\label{ldualgg1}
\end{equation}
It is convenient to rescale the field $V$ defining $V^*V=\frac{e^2}{8\pi^2}$ so
that its
kinetic term has a standard form
\begin{equation}
{\cal L}=\partial_\mu V^*\partial^\mu V+h(V^2 +V^{*2})
\label{ldualgg}
\end{equation}
The residual $Z_2$ symmetry is manifest in the lagrangian eq.(\ref{ldualgg}).
The appearance of the $U_\Phi (1)$ symmetry breaking term
leads to important consequences in the spectrum of the theory.
As opposed to QED the spectrum is not gapless.
Since the coefficient $h$ is small,
the lightest excitation of Lagrangian eq.(\ref{ldualgg}) (when $<V>\ne 0$)
is the phase
of $V$ which has a small mass
\begin{equation}
m= 2\sqrt{h}
\label{mphoton}
\end{equation}

The explicit symmetry breaking causes a more dramatic change
for topologically charged particles.
The theory eq.(\ref{ldualgg}) posesses a topological charge
\begin{equation}
Q=-\frac{i}{4\pi}\int \epsilon_{ij}\partial_i(V^*\partial_j V-c.c.)
\label{q1}
\end{equation}
Excitations that carry this charge are topological solitons of the field
$V(x)$. In as
far as eq.(\ref{ldualgg}) describes the low energy physics of the Georgi -
Glashow model,
these topological solitons must be identified with excitations of the original
theory.
This identification is quite straightforward. The charge $Q$ of eq.(\ref{q1})
can be rewritten in terms of the original variables using eq.(\ref{f1}) and the
connection between $\eta(x)$ and $V(x)$:
\begin{equation}
Q=\frac{e}{32\pi^2}\int \epsilon_{ij}\partial_i\tilde F_j
\end{equation}
and therefore coincides with the topological charge eq.(\ref{qqcd}).
In the Georgi - Glashow model the only
light particles that carry $Q$ are $W^\pm$.
Therefore the lightest topological defects of the field $V$ in
eq.(\ref{ldualgg})
are $W^{\pm}$.
The energy of an isolated defect now is {\it linearly} infrared divergent
rather than
just logarithmically divergent as is the case in QED$_3$.
 The configuration that forms around
the defect is a string rather than a hedgehog (Fig.1). The hedgehog
configuration
is no longer energetically favored since in such a configuration
the phase of the
field $V$ is not equal to that of $<V>\equiv v$ almost everywhere in space.
The energy is
therefore quadratically divergent: $E\propto hv^2 L^2$, where $L$ is
an infrared cutoff. To minimize the energy for a nonzero winding, the system
chooses a stringlike configuration, Fig.1. The phase of $V(x)$
deviates from that of $v$ only within a distance $d$ from the line
stretching from the location of the defect to infinity.

It is easy to estimate the energy of such a static configuration.
The contribution of the
gradient term in Hamiltonian is proportional to $\frac{v^2}{d}$ and
that of the potential to $hv^2dL$. Together
\begin{equation}
E/L=a\frac {v^2}{d}+bhv^2d
\label{E1}
\end{equation}
where $a$ and $b$ are positive numerical factors of order 1. We now
optimize the energy with respect to variation of $d$.
Not surprisingly the optimal width comes out as
\begin{equation}
d\propto \frac {1}{\sqrt{h}}\propto \frac {1}{m}
\label{d}
\end{equation}
and the energy per unit length (the sting tension) is
therefore
\begin{equation}
\sigma \equiv E/L \propto v^2m
\label{sigma}
\end{equation}

The configuration of a defect and an antidefect has the energy proportional to
the
distance between them. This is true if the separation between the defect and
the
antidefect is not too large. When the energy of the string exceeds the energy
needed
for formation of a soliton - antisoliton pair (the core energy) such a pair
will be
formed and the minimal energy state will contain two weakly interacting pairs,
each held
together by a string. This is the same effect which is responsible for
perimeter law
behaviour of very large adjoint Wilson loops in the standard approach.
Nevertheless
as long as the distance between a soliton and an antisoliton is not too large,
the potential is approximately linear.

The quantity $\sigma$ in eq.(\ref{sigma}) pertains to the linear potential
between $W^+$
and $W^-$ and is therefore the adjoint string tension. It is interesting to
note
that the calculation of the fundamental string tension in the framework of the
dual
Lagrangian eq.(\ref{ldualgg}) gives the value of the same order of magnitude.
Let us estimate the vacuum expectation value of the fundamental Wilson loop
$W(C)$.
The easiest way to do that is by using t'Hooft's dual algebra
\begin{eqnarray}
W(C)V(x)W^\dagger (C)&=&-V(x),\ \ x\in S;\\ \nonumber
&=&V(x),\ \ x\not\in S
\end{eqnarray}
Where $S$ is the area enclosed by the curve $C$. Let us choose the vacuum of
the dual
Lagrangian eq.(\ref{ldualgg}) with $<V>=v>0$. Then the state $W(C)|0>$ has the
field
$V(x)$ negative inside $C$ and positive elsewhere. We now approximate the
vacuum
state of the phase of the vortex operator by the vacuum of a free massive field
of mass
 $m$.
\begin{equation}
|0>=\sqrt N {\rm exp}\{-\frac{1}{2}\int {\rm dx dy}\eta(x)G^{-1}(x-y)\eta(y)\}
\end{equation}
with $N$ - the normalisation factor and
$G^{-1}=2v^2\sqrt{-\Box+m^2}\delta^2(x-y)$
 as follows from
eq.(\ref{ldualgg}).
Then
\begin{equation}
W(C)|0>=\sqrt N {\rm exp}\{-\frac{1}{2}\int {\rm dx dy}
[\eta(x)+\phi(x)]G^{-1}(x-y)[\eta(y)+\phi(y)]\}
\end{equation}
with
\begin{eqnarray}
\phi(x)&=&\pi,\ \ x\in S;\\ \nonumber
 & &0,\ \ x\not\in S
\end{eqnarray}
Finally we have
\begin{eqnarray}
<0|W(C)|0>&=&N\int d\eta(x){\rm exp}\{-\frac{1}{2}\int {\rm dx dy}
\eta(x)G^{-1}(x-y)\eta(y)
\\ \nonumber
&+&[\eta(x)+\phi(x)]G^{-1}(x-y)[\eta(y)+\phi(y)])\} \\ \nonumber
&=&\exp \{-\frac{1}{4}\int{\rm dx dy} \phi(x) G^{-1}(x-y)\phi(y)\} \\ \nonumber
&=&\exp\{-\frac{\pi^2}{2}v^2mS{\rm + perimeter \  contribution}\}
\end{eqnarray}
We find therefore for the fundamental Wilson loop the area law with string
tension
of the same order of magnitude as the adjoint string tension eq.(\ref{sigma}).
This
result is fully consistent with the lattice calculations \cite{greensite}.

The rest of the section is devoted to comments on the dual Lagrangian
eq.(\ref{ldualgg}).
\\ i) In this Lagrangian $V(x)$ was assumed to be a ``$\sigma$ - model type''
field:
\begin{equation}
V^*V=\frac{e^2}{8\pi}
\label{sima}
\end{equation}
The effective Lagrangian describes however only long distance properties
of the model. One expects that when higher derivative terms are taken
into account, the ``radial'' part of $V$ becomes dynamical\footnote{This is
reminiscent
of low energy chiral Lagrangians \cite{bando} in which the field
$\sigma$ appears at intermediate energies.}.
Classically, in the $\sigma$ - model Lagrangian
defects are pointlike and their energy is logarithmically divergent in the
ultraviolet
(the core region).

 In the Georgi - Glashow
 model the self energy of $W^\pm$ is finite.
In the dual Lagrangian this can be achieved by relaxing
the constraint eq.(\ref{sima}).
The $\sigma$ - model field is replaced by the unconstrained complex
field $V$ with large mass of the radial component
\begin{equation}
{\cal L}=\partial_\mu V^* \partial^\mu V + h(V^{*2}+V^2)+
 \mu^2 V^*V-\lambda(V^*V)^2
\label{ldual2}
\end{equation}
The coefficients $\mu$ and $\lambda$ must be such that the expectation value
of $V$ is $v^2=\frac{e^2}{8\pi^2}$.
There are now
three different scales in the theory: 1) $m$ - the small mass of the phase of
$V$,
2) $v$ - the VEV of $V$ and 3) $M_\sigma$ -
the large mass of the radial component of $V$.
At energies $E\ll M_\sigma$ the dynamics of the Lagrangian eq. (\ref{ldual2})
is the same as that of the $\sigma$ - model.
At distances of order $1/M_\sigma$ or less the situation has significantly
changed:
$M_\sigma$ serves as an UV cutoff for low energy physics. The size of the
defect
is $\sim 1/M_\sigma$ and its self energy $M_q\sim v^2 {\rm ln}
\frac{M_\sigma}{v}$.
In fact, even within $\sigma$ - model Lagrangian the $\sigma$ particle appears
as a ``bound state'' due to quantum corrections \cite{arefyeva}, so it
does not have to put in by hand. The only difference would be that $M_\sigma$
will depend on the other two parameters. This particular linkage will
be anyway spoiled by higher derivative terms and we therefore prefer to use the
unconstrained phenomenological Lagrangian eq.(\ref{ldual2}). In fact the
appearance of this scalar particle in the effective dual Lagrangian is a
welcome feature. It just represents the scalar Higgs boson which also appears
perturbatively in the spectrum of the Georgi - Glashow model and is not
confined by nonperturbative effects.
\\ ii) We derived
the dual effective Lagrangian in the framework of the dilute monopole gas
approximation.
Let us now discuss what is the natural extension of the picture presented to
the confining region.
Since the Higgs and the confining regions are connected analytically the
spectrum should be qualitatively the same. Of course there are quantitative
differences. The most important one from our point of view is that the lowest
energy excitation in the confinement regime is presumably the scalar
glueball\footnote {Depending on the mass of the matter field it may be more
convenient to think about this state as a scalar meson rather than a scalar
glueball.}
rather than a pseudoscalar "photon".
However it appears to be still possible to describe the low energy dynamics by
the same dual Lagrangian. The reason is that as it stands the Lagrangian
eq.({\ref ldual2}) already contains both, the pseudoscalar and the scalar
degrees of freedom. The phase of the field $V$ is a pseudoscalar. In the Higgs
region it interpolates the massive photon and in the confinement region it
interpolates a pseudoscalar glueball. The radial degree of freedom interpolates
the scalar Higgs particle and a scalar glueball respectively. In order to
describe the confinement regime the coefficients of the Lagrangian must change
so that the level crossing between the scalar and the pseudoscalar excitations
takes place. This is easily achieved if we take in eq.(\ref{ldual2})
$2h+|\mu^2|>0$ and $h>\lambda$. Then the expectation value of the field $V$,
the pseudoscalar mass $M_{\eta}$ and the scalar mass $M_\sigma$ are given by
\begin{equation}
v^2=\frac{2h+\mu^2}{2\lambda}, \  \ M_\eta =8hv^2,  \ \ M_\sigma =8\lambda v^2
\end{equation}
The important point is that as in the Higgs phase $v\ne 0$. It is clear on
general grounds that this should be the case. If $v= 0$ the global $Z_2$
symmetry would be broken. In that case the two regions would be necessarily
separated by a phase transition. However, as mentioned above the phases are
connected analytically and therefore the modes of realization of global
symmetries must be the same.
The topological current $J_\mu$ can
still be expressed in terms of $V$, eq.(\ref{J}). Since $v\ne 0$, the charge
$Q$
is never screened (as in the
Higgs phase) and is well defined everywhere \cite{szlachanyi}.
Classically then the effective dual Lagrangian admits solitonic configurations.
It is very natural to think of these solitons as of constituent quarks in the
same way as in the Higgs phase.
The energetics of these topological defects will be slightly different.
To see this first consider the limit $M_\eta >> M_\sigma$. In this limit it is
clearly energetically unfavorable to perform a wind of the angle $\eta$ at
those points in space where the radial part of $V$ has large value. What will
happen therefore is that at the points where $\eta$ is not constant (inside the
flux tube) the radial part of $V$ will be not close to its expectation value
but rather close to zero. The contribution of the phase $\eta$ to the string
tension will then be negligible. However the radial part will now differ from
its VEV along the string inside which $\eta$ winds. The width of the strip
inside which the radial part of $V$
varies is  given by $1/M_\sigma$. Performing the same rough estimate for the
string tension as in the Higgs regime we find
\begin{equation}
\sigma\propto v^2M_\sigma
\end{equation}
We see therefore that the string tension is dominated by the lightest
excitation which in this case is a scalar glueball.  In the regime where
the scalar and the pseudoscalar masses ar comparable evidently both will
contribute to the string tension.

To conclude this section, although we are unable to derive the dual Lagrangian
directly in the confining region there is a natural extension of the picture of
confinement of constituent quarks. The basic mechanism of confinement is still
the same. The constituents are topological solitons. The string tension now is
mostly due to the variation of the scalar glueball field but the variation of
this field along the flux tube is catalyzed by the necessity to perform a wind
of the phase $\eta$.

\section{Generalization to SU(N).}

In this section we generalize the analysis
to $SU(N)$ gauge theories with adjoint quarks.

In these theories the
gauge $SU(N)$ can be classically broken down to variety of subgroups.
As in the Georgi - Glashow model, classically there are phase transition lines
which
are erased by quantum effects. We will have in mind the region of parameters
in which the dilute monopole gas approximation
is applicable, namely the region in which classically $SU(N)\rightarrow
U(1)^{N-1}$ \cite{snyderman,wadia}. The Lagrangian is
\begin{equation}
{\cal L}=-\frac{1}{2}{\rm Tr}  \ F_{\mu\nu}F^{\mu\nu}
+{\rm Tr}\ |{\cal D}_\mu\phi|^2-U(\phi)
\label{lad}
\end{equation}
where we use matrix notations
($\lambda^a$ are traceless hermitian matrices,
 ${\rm Tr}\lambda^a\lambda^b=\frac{1}{2}\delta^{ab}$)
$$A_\mu\equiv A_\mu^a\lambda^a,\ \ \phi\equiv \phi^a\lambda^a,\ \ $$
\begin{equation}
F_{\mu\nu}=\partial_\mu A_\nu-\partial_\nu A_\mu+ie[A_\mu ,A_\nu],\  \
{\cal D}_\mu\phi\equiv\partial_\mu\phi+ie[A_\mu,\phi]
\end{equation}
The potential $U$ is chosen in such a way that VEV of $\phi$ has $N$
nondegenerate
eigenvalues. There are
$N-1$ (the rank of $SU(N)$ group) mutually commuting generators
which commute with $\phi$. In the basis in which $\phi$
is diagonal these are the diagonal generators
\begin{equation}H_1\equiv \frac {1}{2} diag\{1,-1,0...0\},\ \ H_2\equiv \frac
{1}{2\sqrt{3}}
diag\{1,1,-2,0...0\}, ... ,
\end{equation}
$$H_r\equiv \frac {1}{\sqrt{2r(r+1)}} diag\{1,...,1,-r,0,...0\}, ... ,
H_{N-1}\equiv \frac {1}{\sqrt{2N(N-1)}} diag\{1,...,1,-(N-1)\}$$
These generators define $N-1$ unbroken directions each corresponding to an
``electric''
charge. Therefore any particle in the theory is characterized by the $N-1$
dimensional vector of eigenvalues: ${\vec q}\equiv
(q_1,...,q_{N-1})$. The charges of gauge bosons are roots of $SU(N)$. The $N-1$
zero
roots correspond to (classically) massless
``photons'', while the charges of massive gauge bosons, $W$,
are nonzero roots.
Charged $W$'s are linearly confined, as follows for example from the dilute
monopole
gas approximation.

It is clear that instead of just one topological charge as in the $SU(2)$
case, here one has to define $N-1$
independent topological charges.
The above description is not gauge invariant. We will now
construct explicitly the gauge invariant expression for
the charges.

Let us start by defining $N-1$
(gauge invariant) analogs of $\tilde f_\mu$
eq.(\ref{fqcd}).
They will be defined as
\begin{equation}
\tilde f^\mu_r\equiv 2{\rm Tr} \ \tilde F^\mu \hat \phi_r
\end{equation}
where $N-1$, ($N\times N$) hermitian traceless
matrices $\hat\phi_r$ satisfy the following conditions:
\newline
1) $\hat\phi_r$ transforms under the gauge group covariantly as an adjoint
representation;
\newline
2) $[\hat\phi_r,\hat\phi_s]=0$;
\newline
3) ${\rm Tr}\ \hat \phi_r\hat\phi_s=\frac{1}{2}\delta_{rs}$
\newline
4) The eigenvalue spectrum of $\hat\phi_r$ should be identical to that of $H_r$
for
any $r=1,...,N-1$.

The matrices $\hat\phi_r$ can be constructed from a single matter field $\phi$
in the form
\begin{equation}
\hat\phi_r=\sum_{n=1}^{N-1} \alpha_{rn} \phi^n
\end{equation}
where $\alpha_{rn}$ depend only on the invariants Tr $\phi^m$.
The first three conditions are satisfied by the set $\hat\phi^{'} _r$
constructed using the elementary
Gramm - Schmid algorithm. The only nontrivial condition is 4).
To satisfy it we will work in the basis in which $\hat\phi ' _r$ are diagonal
and
the eigenvalues $x_k$ of $\hat\phi ' _1$ are ordered $x_1>x_2>...>x_N$.
Since $\hat\phi ' _1\propto\phi$ and the eigenvalues of $\phi$ near the vacuum
are
all different, this ordering is always possible. In this basis $\hat\phi ' _r$
have a form
\begin{equation}
\hat\phi ' _r=O_{rs}H_s
\label{O}
\end{equation}
where $O_{rs}$ is an orthogonal matrix that depends only on the invariants.
The equations determining the matrix elements of $O$ are found
taking trace of powers of eq.(\ref{O}).
The solution exists by construction. Having found $O$, we define
\begin{equation}
\hat\phi_r=O_{sr}\hat\phi ' _s
\end{equation}
which satisfies all the requirements.

As an example consider $SU(3)$. In this case
\begin{equation}
\hat\phi ' _1=\frac{\phi}{\sqrt{ 2{\rm Tr}\phi^2}}
,\  \  \hat\phi ' _2=
\frac{1}{\sqrt{2[-2({\rm Tr}\hat\phi ' _1 {}^3)^2+{\rm Tr}\hat\phi ' _1 {}^4]}}
[-2({\rm Tr}\hat\phi ' _1 {}^3)\hat\phi ' _1+\hat\phi ' _1 {}^2]
\label{GS}
\end{equation}
The orthogonal $2\times 2$ matrix $O$ is parametrized
\begin{equation}
O=\left (\begin{array}{ll}
{\rm cos}\theta & {\rm sin }\theta\\
-{\rm sin}\theta & {\rm cos }\theta
\end{array} \right)
\end{equation}
The resulting equation for $\theta$ is
\begin{equation}
{\rm sin}\theta (3-4{\rm sin}^2\theta )=4\sqrt{3}{\rm Tr}\hat\phi ' _1{}^3
\end{equation}
The system of algebraic equations for $N>3$ is in general complicated. We will
not
need its explicit solution but only the knowledge that it exists.

The gauge invariant form for the conserved topological currents $J_\mu^r$ is
\begin{equation}
J_\mu^r=\frac{1}{e}\epsilon_{\mu\nu\lambda}\partial^\nu \tilde f^{r\lambda}
\end{equation}
In order to represent $J_\mu^r$ as manifestly topological we define vortex
operators
 $V_{\vec g}$
\begin{equation}
V_{\vec g}(x)=\exp 2i\sum_r g_r\int d^2y\left[
\frac{1}{e}\epsilon_{ij}{(x-y)_j\over (x-y)^2}
{\rm Tr}\hat\phi_r(y)E_i(y)+\Theta(x-y)J^r_0(y)\right]
\label{Vr}
\end{equation}
Those are again singular gauge transformations with gauge functions
\begin{equation}
\chi_{\vec g}(y)=2\sum_r g_r\Theta( x- y)\hat\phi_r( y)
\end{equation}
An operator $V_{\vec g}$ is local only for values of $g_r$ for which a
discontinuity
of the phase in eq.(\ref{Vr}) is an integer multiple of $2\pi$.
This holds if
\begin{equation}
\sum_r g_rq_r=\frac{k}{2}
\label{diracna}
\end{equation}
 with an integer $k$ for any eigenvalue of $Q_r$ in the theory. This is the
analog of the Dirac quantization condition.

For the theory involving matter
in adjoint representation only, the allowed values of
$Q_r$ lie on the $N-1$ dimensional lattice spanned by the root
vectors of $SU(N)$.
The allowed values of $\vec g$ therefore lie on the dual lattice. For
$SU(N)$ the dual lattice is spanned by weights of the fundamental
representation \cite{snyderman,wadia}. Only $N-1$ operators $V_{\vec g}$
are independent, all the others being equal to products of integer powers
of these $V_i$. For convenience we will always choose such a set $V_i$
which is symmetric under inversion with respect to a hyperplane perpendicular
to the isospin axis (the generator $H_1$).

For concreteness, in the case of $SU(3)$ the nonvanishing roots are
\begin{equation}
\vec r_1=(\frac{1}{2},\frac{\sqrt{3}}{2}),\  \
\vec r_2=(-\frac{1}{2},\frac{\sqrt{3}}{2}),\  \
\vec r_3=(1,0),  \ \ \vec r_4=-\vec r_1, \ \
\vec r_5=-\vec r_2, \ \ \vec r_6=-\vec r_3
\label{rootssu3}
\end{equation}
The fundamental weights are
\begin{equation}
\vec w_1=(\frac{1}{2},\frac{1}{2\sqrt{3}}),\  \
\vec w_2=(-\frac{1}{2},\frac{1}{2\sqrt{3}}),\  \
\vec w_3=(0,-\frac{1}{\sqrt{3}}),\  \
\label{weightssu3}
\end{equation}
The elementary flux operators $V_1$ and $V_2$ correspond to $\vec g_1=\vec w_1$
and
$\vec g_2=\vec w_2$.

The currents $J_\mu^r$ are manifestly topological
when expressed in terms of $V_i$.
\begin{equation}
\vec g_i \vec J_\mu=-\frac{i}{4\pi}\epsilon_{\mu\nu\lambda}\partial^\nu(V_i ^*
\partial^\lambda V_i-c.c.)
\label{Jr}
\end{equation}

We now derive the dual effective Lagrangian for $V_i$.
As in the $SU(2)$ case, the currents $\tilde f_\mu^r$ are not conserved. One
can again
find a modification which is classically conserved
\begin{equation}
\tilde F_\mu^r=\tilde f_\mu^r-\frac{4N(N-1)}{e}\epsilon_{\mu\nu\lambda}
{\rm Tr}\ \hat\phi_r {\cal D}^\nu\hat\phi_r {\cal D}^\lambda\hat\phi_r
\label{Fr}
\end{equation}
(no summation over $r$ is implied.)
The (classical) conservation of $\tilde F_\mu^r$ can be easily
checked\footnote{
This is readily seen in the unitary gauge in which $\hat\phi_r=H_r$.}.
The algebra of the vortex operators $V_r$ and the flux generators $B_r$ is
\begin{equation}
[V_i(x),\vec B(y)]=-{4\pi \over e} \vec g_i V_r(x)\delta^2(x-y)
\label{algebra}
\end{equation}
The flux symmetry in our model on the classical level is $U(1)^{N-1}$. It is
spontaneously broken with $N-1$ massless photons as Goldstone bosons.
Nonperturbative quantum corrections qualitatively change the picture. The
symmetry is
broken explicitly down to a discrete subgroup and the photons as a consequence
become massive. This will be seen directly from the effective Lagrangian
derived in
the dilute monopole gas approximation.

The monopole solutions of this theory were described in
\cite{snyderman,wadia}.
The elementary monopoles have magnetic charge vectors proportional to
the roots of $SU(N)$
\begin{equation}
\vec m=\frac{4\pi}{e}\vec r
\end{equation}
In the dilute monopole gas approximation the ``anomaly equation''
for the magnetic flux currents is
\begin{equation}
\partial^\mu \vec{\tilde F}_\mu (x)=\sum_a {\vec m}_a \delta^3(x-x_a) \ \ \ \ \
\ \
\label{anomalyr}
\end{equation}
Following the standard procedure we introduce into the path integral
fields $\vec\eta$ via
\begin{equation}
\Box\vec\eta=\frac{8\pi}{e}\sum_a \vec m_a\delta^3(x-x_a)
\label{eta}
\end{equation}
In terms of the fields $\eta$
the effective Lagrangian is
\begin{equation}
{\cal L}=\frac{e^2}{32\pi^2}[(\partial_\mu\vec\eta )^2+
\sum_{\alpha}M_\alpha ^2{\rm exp}(i\vec r_\alpha\vec \eta)]
\label{leta}
\end{equation}
where the summation runs over all the $N(N-1)$ nonvanishing roots.
The masses $M_\alpha$ are $M_\alpha ^2\sim \frac{M_W ^{7/2}}{e^3}\exp
(-\frac{4\pi}{e^2}
M_W)$ where $M_W$ is the mass of the gauge boson corresponding to the root
$\vec r_\alpha$.
We now reexpress this Lagrangian in terms of the elementary vortex operators
$V_i$. The simplest way to do this is to relate the fields $\eta_r$
to the phases $\chi_i$ of the vortex operators. From eq.(\ref{algebra})
it follows that
\begin{equation}
[\chi_i(x),\vec B(y)]=\frac{4\pi i}{e}\vec g_i\delta^2(x-y)
\label{chib}
\end{equation}
The commutation relation of $\eta$ and $B$ follows from
eqs.(\ref{anomalyr},\ref{eta}) and the kinetic term in
eq.(\ref{leta}):
\begin{equation}
[\eta_r(x),B_s(y)]=\frac{4\pi i}{e}\delta_{rs}\delta^2(x-y)
\label{etab}
\end{equation}
Therefore
\begin{equation}
\chi_i(x)=\vec g_i\vec\eta(x)
\end{equation}
To invert this relation we use the well known orthogonality relation
between the roots and fundamental weights of $SU(N)$. For given independent
$N-1$ fundamental weights $\vec w_i$ there are $N-1$ roots $\vec r_i$ that
satisfy
\begin{equation}
\vec w_i\vec r_j=\frac{1}{2}\delta_{ij}
\label{wr}
\end{equation}
Hence
\begin{equation}
\vec\eta(x)=2\vec r_i\chi_i(x)
\end{equation}
Substituting this relation into the effective Lagrangian eq.(\ref{leta})
and rescaling the fields $V_i$ we obtain
\begin{equation}
{\cal L}=\partial_\mu V^* _i\partial^\mu V_i+\sum_\alpha h_\alpha
\prod_i V_i ^{2\vec r_\alpha\vec r_i}
\label{lV}
\end{equation}

In particular for $SU(3)$ and $\vec g_i$ and $V_i$ as chosen in
eq.(\ref{weightssu3}) we get
\begin{equation}
\vec r_1=(\frac{1}{2},\frac{\sqrt{3}}{2}),
\  \ \vec r_2=(-\frac{1}{2},\frac{\sqrt{3}}{2})
\label{qi}
\end{equation}
and
\begin{equation}
{\cal L}=\partial_\mu V^* _i\partial^\mu V_i+
 h_1(V_1V^* _2+V^* _1V_2)
+h_2(V_1V^2 _2+V_2V^2 _1 +c.c)
\label{lVsu3}
\end{equation}
The $U(1)\otimes U(1)$ classical flux symmetry is anomalously broken
down to its $Z_3$ subgroup
\begin{equation}
V_1\rightarrow e^{i\frac{2\pi}{3}k}V_1, \  \
V_2\rightarrow e^{i\frac{2\pi}{3}k}V_2
\label{topsym}
\end{equation}
This is an example of
the $Z_N$ ``topological'' symmetry discussed by t'Hooft \cite{thooft1}.
The symmetry $V_i\leftrightarrow V_i ^*$ is the standard charge conjugation.
In addition to it the Lagrangian possesses the $Z_2$ symmetry
$V_1\rightarrow V_2$. This is the charge conjugation symmetry associated
with the charge $Q_1$. It is present in the dual Lagrangian for any $N$.
The reason is that for any root vector $\vec r_\alpha$
of the form $(a,b,...,c)$
there is a counterpart $(-a,b,...,c)$ and the W - bosons corresponding to these
roots are degenerate.

For $SU(4)$ the corresponding results are the following.
The roots and the fundamental weights are
\begin{equation}
\begin{array}{lll}
\vec r_1=(\frac{1}{2},\frac{\sqrt{3}}{2},0)&
\vec r_2=(-\frac{1}{2},\frac{\sqrt{3}}{2},0)&
\vec r_3=(0,-\frac{1}{\sqrt{3}},\sqrt{\frac{2}{3}}) \\
\vec r_4=(1,0,0)&\vec r_5=(\frac{1}{2},\frac{1}{2\sqrt{3}},\sqrt{\frac{2}{3}})
&\vec r_6=(-\frac{1}{2},\frac{1}{2\sqrt{3}},\sqrt{\frac{2}{3}}) \\
\vec r_{i+6}=-\vec r_i
\end{array}
\label{rootssu4}
\end{equation}
\begin{equation}
\vec w_1=(\frac{1}{2},\frac{1}{2\sqrt{3}},\frac{1}{2\sqrt{6}}),\  \
\vec w_2=(-\frac{1}{2},\frac{1}{2\sqrt{3}},\frac{1}{2\sqrt{6}}),\  \
\vec w_3=(0,0,\frac{3}{2\sqrt{6}}),\  \
\vec w_3=(0,-\frac{1}{\sqrt{3}},\frac{1}{2\sqrt{6}}),\  \
\label{weightssu4}
\end{equation}
The elementary vortex operators are chosen to correspond to $\vec w_1$,
 $\vec w_2$ and $\vec w_3$. The dual Lagrangian is
$$
{\cal L}=\partial_\mu V^* _i\partial_\mu V_i+
 h_1(V_1V^* _2+V^* _1V_2)+ h_2(V_1V^* _3+V_2V^* _3 +c.c.)+
$$
\begin{equation}
h_3(V_1V^2 _2V_3+V^2 _1V_2V_3 +c.c)+h_4(V_3^2V_1V_2+c.c)
\label{lVsu4}
\end{equation}
It again displays t'Hooft's $Z_4$ symmetry and both charge conjugations.
It is clear that these symmetries remain in the dual Lagrangian for any $N$.

The dual Lagrangian eq.(\ref{lV}) exhibits the same confinement mechanism as
the
Georgi - Glashow model. When the fields $V_i$ acquire
nonzero VEV, the solitons are linearly confined. In particular, charged
$W$'s are confined into neutral (with respect to all topological charges)
composites. In the $SU(3)$ the gauge bosons corresponding
to the roots $\vec r_1,$ $\vec r_2$ and $\vec r_3$ are a defect of $V_1$, a
defect of
$V_2$ and a bound pair of $V_1$ - defect with a $V_2$ - antidefect
(as follows from eq.(\ref{Jr})).

\section{Baryons.}

The nonabelian gauge theories with quarks in adjoint
representation of the color group
are the simplest
models exhibiting the topological mechanism
of confinement. The most important phenomenon present in QCD
but absent in the Georgi - Glashow model is the appearance of baryons. The
baryon
number is defined only when there are quarks belonging to the fundamental
(or other nonzero $N$ - ality)  representation of the color group.

The mesons in the dual representation should be again configurations
of defect - antidefect connected by a string. As for a baryon, clearly it
should contain
$N$ defects each carrying $1/N$ fraction of baryon number (constituent
quark). The topological charge should be neutralized by an (baryon number zero)
antidefect with topological charge $-N$. This will require the existence of two
types of
defects in the theory. We will see that this is indeed the case.
The $SU(3)$ baryon in this picture is a
{\bf Y}-shape configuration \cite{thooft2}.

We now consider the model which in addition to adjoint fields contains
a fundamental multiplet of scalar quarks $\theta^A$.
\begin{equation}
{\cal L}=-\frac{1}{2} {\rm Tr} \ F_{\mu\nu}F^{\mu\nu}+{\rm Tr}\ |{\cal
D}_\mu\phi|^2
+|{\cal D}_\mu^{AB}\theta^B|^2-U(\phi,\theta)
\label{lfundamental}
\end{equation}

The dual Lagrangian still contains vortex operators, but the elementary
vortex operators now are different. The flux currents are defined by
eq.(\ref{Fr}).
A vortex operator again has a form of eq.(\ref{Vr}) but the locality is
retained
only for a subset of vectors $\vec g$. Since the theory now contains
fundamental
charges the Dirac quantization condition eq.(\ref{diracna}) must be satisfied
with $\vec q$ being weights of
the fundamental  as well as the adjoint representations. As a result $\vec g$
lie on the lattice spanned by roots of $SU(N)$. The $N-1$ elementary
local operators $V_i$ correspond to $N-1$ independent root vectors. For example
in $SU(3)$ we choose $\vec g_1=\vec r_1$ and $\vec g_2=\vec r_2$ of
eq.(\ref{rootssu3}).
The dilute monopole gas approximation applied to the present theory gives
the same effective Lagrangian eq.(\ref{leta}) in terms of the fields $\vec
\eta$.
The relation between $\vec \eta$ and $\chi_i$ is modified since $V_i$ now carry
different values of magnetic fluxes. In terms of the new $V_i$ we obtain
\begin{equation}
{\cal L}=\partial_\mu V^* _i\partial^\mu V_i+\sum_\alpha h_\alpha
\prod_i V_i ^{2\vec r_\alpha\vec w_i}
\label{lVfund}
\end{equation}

For $SU(3)$
\begin{equation}
{\cal L}=\partial_\mu V^* _i\partial^\mu V_i+
h_1(V_1V^* _2+V^* _1V_2)
+h_2(V_1+V_2 +c.c)
\label{lVfundsu3}
\end{equation}
The main difference between eq.(\ref{lV}) and eq.(\ref{lVfund}) is that in
the latter no residual flux symmetry remains. This however, does not
change the qualitative picture of confinement.

The derivation is valid in the region of parameter space where the phases of
all
$N-1$ vortex operators are light variables. The mechanism of confinement
remains
essentially the same
in the case when one of the variables is much lighter
than the others. In this case the
low energy effective dual
Lagrangian should contain only the light vortex field $V$. The topological
charges that are carried by the defects of heavy $V_i$ are confined at small
distances.
The only topological charge relevant at low energies is the one that
corresponds
to the light $V$.
This situation is realized  in the region of parameters in which classically
the $SU(N)$ gauge symmetry is broken down to $ U(1)$. From now on
we will restrict ourselves to this type of models.

Our goal now is to
investigate how to incorporate the baryon number in the dual Lagrangian.
Nontopological global symmetries should be represented
linearly. This was the case with both charge conjugations in eq.(\ref{lV})
and with flavor symmetries in abelian gauge theories \cite{mpl}.
The vortex field $V$ is invariant under baryon number transformations.
At the energy scale where baryons appear, one cannot
describe physics by effective Lagrangian involving $V$ alone. We introduce
therefore a baryon number charged field $W$.

To this end
we take a ``phenomenological'' approach. Interactions of the fields $V$ and
$W$ should be such
that the elementary defect of the field $V$ carries baryon number $1/N$. This
soliton will represent the constituent (fundamental) quark in our string
picture.
 A necessary condition
is that in the core of the defect (where $V$ vanishes) the field $W$ has a
nonzero value.
Outside the core, $W$ approaches its VEV and thus
vanishes (since the baryon number symmetry is not broken
spontaneously). The interaction potential
 should favor the configurations in which for $V$ close to
its VEV, $W$ is small whereas for small $V$, $W$ is nonzero. This is
conveniently achieved imposing the constraint\footnote{The same effect
can be achieved by adding an interaction term
$\lambda (V^*V+W^*W-\frac{e^2}{8\pi^2})^2$. However for long
distance properties which we discuss these two alternatives are
indistinguishable. We choose the $\sigma$ -
model type constraint to simplify the calculations.}
\begin{equation}
V^*V+W^*W=\frac{e^2}{8\pi^2}
\label{constr}
\end{equation}

In order to attach the baryon number to the topological soliton, three
derivative terms are
necessary. The following Lagrangian satisfies the above
requirements
\begin{equation}
{\cal L}=\partial_\mu V^*\partial^\mu V
+\partial_\mu W^*\partial^\mu W +h(V^*+V)+
\frac{1}{2\pi N}\epsilon^{\mu\nu\lambda} w^*\partial_\mu w
\partial_\nu( v^*\partial_\lambda v)
\label{mthreeder}
\end{equation}
where $w\equiv{W\over \sqrt{W^*W}}, v\equiv {V\over\sqrt{V^*V}}$.
The linear term drives the VEV of $V$ to its maximal value.
Consequently due to eq.(\ref{constr}) the VEV of $W$ vanishes.

We calculate the quantum numbers of the low lying solitons using
the zero mode quantization method \cite{zahed}.
Classically there are degenerate static solution of the equations of motion.
Let us start with the elementary solitons (a single wind of $V$). In fact,
there is a one parameter set of degenerate classical solutions. The solutions
\begin{equation}
V(x)=v_0(x),\    \ W(x)=w_0(x) e^{i\alpha}
\label{solitons}
\end{equation}
have the same energy for any $0<\alpha<2\pi$ due to the baryon number symmetry.
The zero mode $\alpha$ is a slow variable,
 the only one that has to be treated quantum mechanically\cite{zahed}.
 Substituting the solution eq.(\ref{solitons}) into the Lagrangian we get
\begin{equation}
{\cal L}=\frac {1}{2}\mu \dot\alpha^2-\frac {1}{N} \dot\alpha+const
\label{zerol}
\end{equation}
where
\begin{equation}
\mu\equiv 2\int d^2x w^*_0(x) w_0(x)
\label{mu}
\end{equation}

The momentum conjugate to $\alpha$ is
\begin{equation}
p=\mu \dot\alpha-\frac {1}{N}
\label{p}
\end{equation}
and the corresponding Hamiltonian:
\begin{equation}
H=\frac {1}{2\mu} (p+\frac {1}{N})^2
\label{H}
\end{equation}
Since $\alpha$ is an angular variable, $p$ has only integer eigenvalues.

We first consider the generic case $N>2$. In this case
the vacuum of the one soliton sector is nondegenerate: $p=0$.
The baryon number current is
\begin{equation}
J^B_\mu=i(W\partial_\mu W^*-W^*\partial_\mu W)
\label{J^B}
\end{equation}
In this sector the baryon number is
\begin{equation}
Q^B=\mu\dot\alpha=p+\frac {1}{N}
\label{baryonn}
\end{equation}
Therefore the elementary soliton has baryon number $1/N$.

The case $N=2$ is special. The vacuum of the Hamiltonian eq.(\ref{H})
is doubly degenerate: $p=0$ and $p=-1$. The two solitons have baryon numbers
$+1/2$
and $-1/2$ respectively.
This degeneracy corresponds to
 an additional global
$SU(2)$ symmetry
 of the $SU(2)$ gauge model eq.(\ref{lfundamental})
 - the ``custodial'' symmetry:
\begin{equation}
\theta\rightarrow\theta U
\label{custodial}
\end{equation}
where $\theta $ is a $2\times 2$ matrix:
\begin{equation}
\theta\equiv
\left (\begin{array}{ll}
\theta_1 & \theta_2^*\\
\theta_2 & -\theta_1^*
\end{array} \right)
\label{}
\end{equation}
Baryon number is one of the $SU(2)$ generators. Therefore particles always come
in pairs
with opposite baryon number\footnote{To represent faithfully the custodial
symmetry in the low energy Lagrangian one should promote the field $W$
to an $SU(2)$ multiplet. This is done in Appendix.}.

Now we return to the generic case $N>2$. The elementary soliton with baryon
number $1/N$ evidently represents a constituent quark. However, as we discussed
in the
beginning of this section, a consistent picture of baryons require the
existence
of another elementary soliton with quantum numbers $Q=N, Q^B=0$. Let us
systematically
investigate the quantum numbers in different soliton sectors.

In the sector with $Q=q$, the zero mode Lagrangian is
\begin{equation}
{\cal L}=\frac {1}{2}\mu_q \dot\alpha^2-\frac {q}{N} \dot\alpha+const
\label{0l}
\end{equation}
The momentum conjugate to $\alpha$ is now
\begin{equation}
p=\mu_q \dot\alpha-\frac {q}{N}
\label{pq}
\end{equation}
The corresponding Hamiltonian is:
\begin{equation}
H=\frac {1}{2\mu_q} (p+\frac {q}{N})^2
\label{Hq}
\end{equation}
For $q/N<1/2$ the vacuum is $p=0$, while for $1/2<q/N<1$ the vacuum is $p=1$.
The baryon number is
\begin{equation}
Q^B=\mu_q \dot\alpha=p+\frac {q}{N}=
\left\{ \begin{array}{lll}
\frac {q}{N}\ \ {\rm for}\ \ q<\frac {N}{2} \\
\pm \frac {1}{2} \ \ {\rm for} \ \ q=\frac {N}{2} \\
-1+\frac {q}{N} \ \ {\rm for} \ \ \frac {N}{2}<q<N
\end{array}\right.
\label{baryon}
\end{equation}

Remarkably, for $q=N$ the soliton has baryon number zero. Moreover this is the
only
additional independent constituent one needs. We will call it ``constituent
gluon''
since its quantum numbers coincide with those of gauge bosons. All the
others can be thought of as composites of the constituent quarks and gluons.

For $SU(3)$ and $SU(4)$ these are given in Tables 1 and 2\footnote{Note the
degeneracy at $q=N/2$ for any even $N$. Presumably more accurate treatment
lifts this degeneracy since there is no symmetry in eq.(\ref{mthreeder})
accounting
for it.}.

\begin{table}
\begin{center}
\begin{tabular}{||c|c|c||}
\hline \hline
$q$ & $B$ & composition \\ \hline  \hline
1 &$\frac {1}{3}$ &$q$ \\ \hline
2 & $-\frac {1}{3}$ &$\bar q g$ \\ \hline
3 & 0 & $g$ \\ \hline
\end{tabular}
\end{center}
\label{table1}
\caption{Quantum numbers of solitons for the $SU(3)$ gauge theory.}
\end{table}

\begin{table}
\begin{center}
\begin{tabular}{||c|c|c||}
\hline \hline
$q$ & $B$ & composition \\ \hline  \hline
1 &$\frac {1}{4}$ &$q$ \\ \hline
2 & $\pm\frac {1}{2}$ &$qq,\bar q \bar q g$ \\ \hline
3 &$-\frac {1}{4}$ &$\bar q g$ \\ \hline
4 & 0 & $g$ \\ \hline
\end{tabular}
\end{center}
\label{table2}
\caption{Quantum numbers of solitons for the $SU(4)$ gauge theory.}
\end{table}

It is interesting to note that the Lagrangian for the theory involving
fundamental quarks has the model considered in sections 2 and 3 as its limit.
One expects that the mass of the field $W$ becomes large together with the mass
of fundamental quarks. Indeed, as the mass of $W$ increases, the region inside
the soliton's core in which $W\ne 0$ shrinks. Consequently $\mu$ in
eq.(\ref{mu})
decreases as well. From the Hamiltonian eq.(\ref{Hq})
it follows that only for $q=N$ (the gluon) the soliton's energy does not
diverge in the limit $m_q\rightarrow 0$. In this limit $m_W\rightarrow\infty$.
Therefore the only ``survivors''
are the solitons with winding $N$ - the constituent gluons. The field
$V'\equiv V^{1/N}$ becomes local and in terms of this field we return to the
$Z_N$ symmetric Lagrangian.

Flavor symmetries can be taken into account in a way similar to that used for
baryon number.

\section{Discussion.}

In this paper we proposed a new qualitative picture of confinement in 2+1
dimensional nonabelian gauge theories. The main idea is that constituent
quarks are topological defects of a local field $V$. This field $V$
was explicitly constructed in terms of original Yang - Mills
variables. In the nonabelian gauge theories the magnetic flux symmetry
$V\rightarrow e^{i\alpha}V$
 is explicitly broken and in the confining phase $<V>=v\ne 0$.
In this situation strings of electric flux are formed between
point defects. The topological charge carried by quarks is thereby
linearly confined. Baryons in this picture contain $N$ constituent
quarks and one ``constituent gluon''and for $SU(3)$ have a {\bf Y}-
shaped form. In the $SU(N)$ gauge theories we identified $N-1$
independent topological charges and $N-1$ corresponding vortex
fields $V_i$.

The vortex operators $V_i$ are closely related to the ``disorder parameter''
introduced in this context by t'Hooft \cite{thooft1}. It is important
to realize, however, that the basic philosophy of t'Hooft's approach is
different from ours. t'Hooft proposed to understand the linear confinement
as formation of line defects, the Bloch walls, between the different vacua
that correspond to different directions of spontaneous breakdown of the
residual $Z_N$ flux symmetry. He showed that in terms of the elementary
vortex fields {\it defined in a theory without fundamentals}
the wall is attached to any fundamental quark. The linear potential is
understood
as constant linear energy density stored in this line defect.

There are two major problems with this idea. First, according to t'Hooft,
there should be no confinement of adjoint quarks. This contradicts
more recent results \cite{greensite} which show that
there is no basic difference between the adjoint and the
fundamental string tension. The picture discussed in this paper does
not suffer from this problem since it indeed leads to the fundamental and the
adjoint string tensions of the same order of magnitude, as shown in Section 2.
The second problem with t'Hooft's approach
is a certain lack of consistency of the picture
itself even for fundamental quarks. It is a mystery where does the Bloch
wall end inside a meson. Obviously, the line defect of this kind should
either stretch all the way to infinity or form a closed ring.
In this picture of a meson, a piece of the line defect
is seen clearly, but the rest (including the location of the quarks)
is blurred by heavy mist. Indeed, it is impossible to visualize a meson
as a well defined configuration of the field $V$, since $V$ defined by t'Hooft,
is not a local
field in a theory with fundamental quarks.

The difference between t'Hooft's approach and the one advocated in this paper
is that t'Hooft emphasizes the importance of line defects (strings)
while we concentrate
on the picture of quarks (both fundamental and adjoint) as point defects.
In a theory with fundamentals the elementary local vortex field
$V_{fundamental}$
carries $N$ units of flux compared to $V_{adjoint}$ in a theory
with only adjoint charges. Therefore in our picture the string is not a line
defect.
It is just a stringlike field configuration which minimizes the energy
of a defect - antidefect pair and thus
necessarily forms when the flux
symmetry group is not a continuous one\footnote{The residual flux symmetry
could be $Z_N$
or could be absent at all, this is not crucial for the phenomenon
of linear confinement in this picture.}.

It is generally believed that confinement in 2+1 dimensions
is quite similar to that in realistic 3+1 dimensional nonabelian
gauge theories like QCD. Some of the key elements of the above
picture can be carried over to 3+1 dimensions.
First, the definition of the topological charge is still given by
eq.(\ref{qqcd}).
As in 2+1, the analogy with the confining phase of 3+1 dimensional compact
QED suggests that this is the correct identification of the confined charge.
It would be very interesting to measure the corresponding form factor
of hadrons directly by Monte Carlo simulations of the lattice gauge theories.
If the identification is correct, the mesons should have large topological
dipole moments.

The flux
symmetry (which in 3+1 we prefer to call ``magnetic symmetry'')
has also a straightforward generalization. This was
done in \cite{ijmp} for noncompact QED. The conserved current is the
dual field strength $\tilde F_{\mu\nu}=\epsilon_{\mu\nu\lambda\rho}
\partial^\lambda A^\rho$ \footnote{The magnetic symmetry defined here should
not be confused with the symmetry generated by the monopole charge. The latter
in QED can be only thought of as a gauge symmetry, since the magnetic monopole
charge density must vanish. In contrast, the magnetic symmetry we are talking
about is a global symmetry. Its generators are magnetic fluxes via any infinite
two dimensional surface and the charge density is the magnetic field. For
discussion see \cite{ijmp}.}. The fact that the current is an antisymmetric
tensor results in nontrivial Lorentz properties
of the photon - the Goldstone
particle appearing due to its spontaneous breaking.

In the Georgi - Glashow model the analog is
\begin{equation}
\tilde F_{\mu\nu}=\tilde
F_{\mu\nu}^a\hat\phi^a-\frac{1}{e}\epsilon_{\mu\nu\lambda\rho}
\epsilon^{abc}
\hat\phi_a ({\cal D}^\nu \hat\phi)^b({\cal D}^\lambda \hat\phi)^c
\label{com4}
\end{equation}
Classically this current is conserved but quantum mechanically it is
broken in the presence of t'Hooft - Polyakov monopoles
\begin{equation}
\partial^\nu \tilde F_{\mu\nu}=J^m_\mu
\label{anom4}
\end{equation}
One can think about this anomaly in the same way we did in 2+1
dimensions. It is natural to assume that a theory may contain a massless boson
only in the presence of a spontaneously broken (global) continuous symmetry.
This ``reverse Goldstone theorem'' has not been proven. Nevertheless in all
known
cases (in $d>1+1$) whenever a massless boson is present, it has been possible
to find a spontaneously broken global symmetry. Massless photons in $QED_4$
also conform to this rule. Assuming for a moment that this folk theorem is
true,
one is lead to interesting predictions for four dimensional gauge theories.
Once the magnetic flux symmetry is explicitly broken, the gauge bosons
can not be massless.
Furthermore, according to Swieca's theorem \cite{swieca}
if there is no massless boson, the charges must be confined (or screened).

 It is well
known that this is the case in confining phases of both, compact QED
and Georgi - Glashow models. In this phase, according to the dual
superconductor picture \cite{thooft2,banks},
the monopoles condense in the vacuum. The anomalous right hand side
of eq.(\ref{anom4}) is therefore large in the vacuum. The effects of the
anomaly such as finite mass of gauge bosons and finite string tension
can be seen summing over monopole loops in the partition function
\cite{polyakov3}. The same calculation in the Coulomb phase does not reveal
confinement of charges. In this phase the monopoles do not condense.
The magnetic symmetry breaking effects are much smaller. However, according to
the above naive argument, they should
manifest themselves in processes involving creation of virtual
monopole - antimonopole pairs. Since in the weak coupling regime the
monopoles are very heavy, the probability of these processes is tiny and the
resulting string tension is expected to be small.

At this point one should make a delicate distinction between the compact
QED (without matter fields) and the Georgi - Glashow model. In QED
without matter fields, although $\tilde F_{\mu\nu}$ is anomalous, the
field strength $F_{\mu\nu}$ is conserved. In fact the theory is dual to
the noncompact QED with electric charges \cite{polley}.
In this case, since there still is a conserved
tensor current, $F_{\mu\nu}$, the theory can contain a massless photon.
However, since in this theory the broken current is a tensor rather than
a pseudotensor, the photon is a pseudovector (unlike in noncompact
QED)\footnote{Although
parity can be redefined to make the photon a vector, we mean the standard
definition of parity in terms of the original fields $A_\mu$.}.

The Georgi - Glashow model on the other hand contains both, magnetic
monopoles and electric charges, so that neither $\tilde F_{\mu\nu}$
nor $F_{\mu\nu}$ is conserved. According to this point of view,
in the Higgs phase of the Georgi - Glashow model the photon should
have a small mass and the electric charges ($W^\pm$) should be
very weakly confined (or screened)\footnote{There is a remote
possibility to define a
combination of $\tilde F$ and $F$ which is conserved.
This is very unlikely since the values of electric and
magnetic charges carried by any particle in the theory are in this case not
constrained to have a fixed ratio. Even if it is possible to redefine
$\tilde F_{\mu\nu}$,
the photon will not have definite parity
and cannot be identified with the massless photon of the perturbation
theory.}. The argument can be made in any nonabelian gauge theory
with semisimple gauge group such as GUTs.
In fact, Swieca's theorem can be applied to magnetic as well as to electric
charges.
Therefore in any phase of the theory the possibilities open to electric {\it
or} magnetic
charges are either to be screened or confined.

This contradicts the widely held belief (inspired by perturbation theory)
that in the Higgs phase of the Georgi - Glashow
model the photon is
massless and $W^\pm$ are free.
However it is not entirely out of the question that the perturbative picture of
the
Higgs - confinement phase transition is inaccurate. To understand the reason,
let us
consider QED rather than QCD. There the Higgs - Coulomb phase transition is
usually
associated with the spontaneous breaking of the electric charge. However the
electric
charge being topological, does not have a
local order parameter and the difference between
the two phases is a nonvanishing expectation value of a {\it nonlocal} charged
operator
in the Higgs phase \cite{szlachanyi}. But in the absence of a local order
parameter
the Goldstone theorem is not applicable and there is nothing that tells us that
the two phases are not analytically connected. In QED this does not happen
since
there is another symmetry (the magnetic symmetry) that is implemented
differently in the two phases. However in the $Z_2$ gauge theory, for example,
which does not have a conserved magnetic flux, the Higgs
and the confining phase are indeed connected analytically \cite{szlachanyi}.
The situation in nonabelian gauge theories may be similar in this respect to a
$Z_2$
gauge theory, in which case one would indeed expect linear confinement of
$W^\pm$.

We stress again that our discussion hinges on the validity of the ``reverse
Goldstone
theorem'' which has not been rigorously proven.
The lattice results on the Georgi - Glashow model
 at this stage can not rule out this possibility. The photon mass in the Higgs
phase was measured in \cite{lattice} and was found to be consistent with zero.
On the other hand the same authors measured the divergence of the magnetic
field and found that it also vanishes. This implies that monopoles are
unimportant in the region of parameters where the calculation was performed
at all scales between the lattice spacing and the size of the lattice. In this
case one would indeed expect that both the anomaly of the magnetic symmetry and
the photon mass will be undetectable.

We note that these conclusions would not apply to the standard model, which
has a gauge group with an invariant abelian subgroup and has therefore a
precisely conserved current $\tilde F$.
Unfortunately there are no adequate analytical tools available to calculate
the photon's mass in theories involving both electric and magnetic charges
\cite{blagojevic}. It would be very interesting to confirm or refute the
conclusions
based on the present simple minded line of reasoning by
direct Monte Carlo
measurement of the photon's mass and the string tension
{\it in the Higgs phase}
of a 3+1 dimensional nonabelian gauge theory like the Georgi - Glashow model.

{\bf Acknowledgements.}
We thank G. Gat, E.G. Klepfish and A. Krasnitz for discussions and M. Marcu
for correspondence. This work is supported by NSERC of Canada.

\section{ Appendix. A Dual Lagrangian with Custodial SU(2) Symmetry.}

In this appendix we construct the low energy dual Lagrangian for the SU(2)
gauge theory with fundamental quarks. In addition to the vortex field
$V$ the custodial symmetry forces us to introduce the complex doublet
$W_i$, $i=1,2$. As in the SU(N) case we still retain the $\sigma$ - model
constraint
\begin{equation}
V^*V+W_i^*W_i=\frac{e^2}{8\pi}
\end{equation}
The standard kinetic term for $W_i$ has larger symmetry than required:
$SO(4)$ rather than $SU(2)$. This symmetry should be reduced by adding
another two derivative term of the form
\begin{equation}
f(W^*W)(W^*\partial_\mu W-W\partial_\mu W^*)^2
\label{twoder}
\end{equation}
This term has the symmetry $U(2)$, which is still too large. However if the
function $f$ is chosen as
\begin{equation}
f(W^*W)=\frac{1}{4(W^*W)}
\end{equation}
the invariant $U(1)$
subgroup of the $U(2)$, the common phase rotation of $W_i$, becomes
gauge symmetry. As a result only the $SU(2)$ group is the global
symmetry on the physical Hilbert space.
To get quarks with correct quantum numbers we also have to
add the (U(1) gauge invariant) three derivative term.
We thus arrive at the dual Lagrangian
$$
{\cal L}=
\partial_\mu V^*\partial^\mu V + \partial_\mu W^*\partial^\mu W
+ \frac{1}{4(W^*W)} (W^*\partial_\mu W-W\partial_\mu W^*)^2
+h(V+V^*)-$$
\begin{equation}
\frac{1}{8\pi (W^*W) (V^*V)}\epsilon^{\mu\nu\lambda}
(W^*\partial_\mu W-W\partial_\mu W^*)\partial_\nu
(V^*\partial_\lambda V-V\partial_\lambda V^*)
\label{lapp}
\end{equation}

In the sector with topological charge $q$ (the interesting values
are $q=1,2$) the degenerate static solutions are of the form
\begin{equation}
V(x)=v^q(x),\  \ W_i(x)=u_{ij}w_i^q(x)
\label{solapp}
\end{equation}
Here $v^q(x)$, $w^q(x)=\delta_{i1}w^q(x)$ is
a particular solution and $u_{ij}$ is a constant unitary matrix.
Substituting eq.(\ref{solapp}) into the Lagrangian eq.(\ref{lapp})
we get the Lagrangian for zero modes $c_i$ (the first column vector of
the matrix $u$)
\begin{equation}
{\cal L}=\mu_q [\dot c\dot c^*+ \frac{1}{4} (c^*\dot c-c\dot c^*)^2]
-\frac{iq}{2} (c^*\dot c-c\dot c^*)
\label{lzeroapp}
\end{equation}

This leads to the following Hamiltonian
\begin{equation}
H=\frac{1}{\mu_q}J^2
\label{hzeroapp}
\end{equation}
where $J^2$ is the quadratic Casimir of the $SU(2)$ group. The gauge
invariance of the lagrangian eq.(\ref{lzeroapp}) also results
in the constraint
\begin{equation}
i(\pi^i c^i-\pi^{*i}c^{*i})=q
\label{conapp}
\end{equation}
 For $q=1$ the constraint allows only representations with duality 1. The
lightest of those according to the Hamiltonian eq.(\ref{hzeroapp})
is the fundamental representation. Thus we get the custodial doublet
of quarks as the lowest lying soliton with $q=1$. For $q=2$ the duality
must be zero and the lowest excitation is the custodial singlet. This is
the gluon (W boson) of the original theory.

\newpage

\begin{figure}
\vspace{8cm}
\caption{A stringlike
configuration of $V$ that is formed around a pointlike defect.}
\end{figure}

\end{document}